\documentclass[12pt]{iopart}
\input axodraw.sty
\def\lsim{\:\raisebox{-0.5ex}{$\stackrel{\textstyle<}{\sim}$}\:}

\begin{document}

\vspace*{-7mm}

\hfill MZ-TH/98-56 

\hfill hep-ph/9901287

\title[HERA Physics Beyond the Standard Model]{HERA Physics Beyond the 
  Standard Model}

\author{H Spiesberger} 

\address{Institut f\"ur Physik, Johannes-Gutenberg-Universit\"at,
  D-55099 Mainz, Germany}

\begin{abstract}
  The prospects of physics beyond the standard model in deep inelastic
  scattering are reviewed, emphasizing some scenarios which attained
  attention after the observation of an excess of events with large
  momentum transfer at HERA\footnote[1]{Talk presented at the 3rd UK
    Phenomenology Workshop on HERA Physics, September 1998, Durham}.
\end{abstract}





\section{Introduction}

After 6 years of running and major improvements of the electron proton
collider machine, the experiments at HERA start to open a new focus of
physics analyses looking at processes with cross sections of the order
of 1 pb and below. This is the typical value for cross sections at large
values of Bjorken $x$ and momentum transfer $Q^2$, or more generally of
processes with large transverse momenta. At the upper limit of the
available center-of-mass energy, the cross sections for deep inelastic
scattering are equally determined by both electromagnetic and purely
weak interactions. Moreover, measurements are possible of rare standard
model processes like the production of an additional gauge boson, or of
radiative processes in neutral and charged current scattering.  Other
examples are the production of lepton pairs or multi-jet systems with
large invariant masses.  These low cross section processes provide a
wealth of possibilities to look for deviations from the standard model
predictions and constitute important backgrounds for searches for
physics beyond the standard model \cite{future,WG2}.

New physics may be found in a search for processes which are forbidden
in the standard model or have tiny cross sections much below the level
of 1 pb; some models of physics beyond the standard model predict such
``gold-plated'', background-free signatures.  However, more common is
the situation that the cross section for conventional standard model
processes are only slightly modified. The first task when searching for
new physics is therefore to obtain a precise and detailed knowledge of
standard model predictions.  For deep inelastic scattering this task is
twofold: on the one hand one has to provide precise parametrizations of
parton distribution functions evolved in $Q^2$ according to
next-to-leading order of QCD. On the other hand, the cross sections for
hard lepton-quark and lepton-gluon subprocesses have to be known also at
least to next-to-leading order. The theoretical tools needed to solve
the first part of the task are well-established and the precision of
parton distribution functions depends mainly on the quality of
experimental data \cite{WG1}. On the other hand, NLO calculations for
hard subprocesses, while continuously improving, have not yet reached a
completely satisfactory status \cite{wkhs}.

NLO calculations for inclusive scattering, i.e.\ $O(\alpha_s)$
corrections to the structure functions ($F_L$, $\Delta F_3$ and, 
in the $\overline{\rm MS}$ scheme, $\Delta F_2$) are known since long
\cite{disnlo}. Considerable progress has been made with respect to jet
production in DIS and photoproduction \cite{WG3}, but $Z$ and $W$
exchange, relevant at large $Q^2$, is not taken into account in the
available NLO Monte Carlo programs.  Only recently, next-to-leading
order corrections to the $W$ production $ep \rightarrow W+X$, including
resolved contributions to photoproduction, have been reported
\cite{spira}. Calculations for NLO corrections to the production of
isolated photons are available for deep inelastic scattering
\cite{isogamma1} and for photoproduction \cite{isogamma2} (see also
\cite{pbmf}), but the transition region of small and large $Q^2$ has not
yet been investigated. For other cases, like for example lepton pair
production, NLO, sometimes even LO, calculations are still missing.

The demands for the precision of standard model predictions varies
depending on the size of the cross section. For inclusive measurements
in deep inelastic scattering one requires rather precise measurements
when searching for new physics. Specific final states, in particular
when they contain one or two particles with high transverse energy, can
be left at a precision of $O(10\,\%)$. For ``generic searches'', where
deviations from standard model predictions are looked for without
referring to any specific expectations \cite{wkhs}, a classification of
``interesting'' final states containing a high-$p_T$ charged lepton,
large missing transverse momentum, or a high-$p_T$ jet, or any
combination of them, would be helpful for a systematic investigation of
experimental uncertainties.

The motivation to search for new physics at HERA has received a strong
impetus by the observation of enhancements of cross sections at several
places. The excess of events at large $x$ and large $Q^2$ in neutral and
charged current scattering \cite{excess} has been discussed at lenght in
the literature, see \cite{altarelli,hsrr} and references therein.
Notably the occurence of events with an isolated muon and large missing
transverse momentum at H1 \cite{h1muon} which are seemingly not a sign
of $W$ production presents a challenge for the understanding of the
experiments. Also at other experiments data suggest themselves as an
indication for the appearance of new physics. I would like to mention
only the high-$E_T$ dijets at Tevatron \cite{dijets} and the cross
section for $e^+e^- \rightarrow W^+W^-$ at LEP2 which is too low at the
highest $\sqrt{s}$ \cite{lep2ww}. However, in the latter case, the
experiments find a suppression, not an enhancement, and it seems more
difficult to find a scenario which predicts the necessary amount of
interference with the standard model.  It is therefore more wide-spread
to believe that this is a statistical fluctuation, but the same can be
the origin of all the other observations as well. Probably the most
strong hint for the presence of new physics is the experimental evidence
for neutrino oscillations \cite{superK}.

In the following, I selected some of the alternatives to standard model
physics which, if realized in nature, have a good chance to be
discovered at HERA. If not, HERA is expected to significantly contribute
to setting limits on their respective model parameters.  Other topics of
interest are discussed in \cite{future,WG2,schrempp}.


\section{The main alternatives}

Despite of the great success of the standard model, various conceptual
problems provide a strong motivation to look for extensions and
alternatives. Two main classes of frameworks can be identified among the
many new physics scenarios discussed in the literature:
\begin{itemize}
\item Parametrizations of more general interaction terms in the
  Lagrangian like contact interactions or anomalous couplings of gauge
  bosons are helpful in order to {\it quantify the agreement} of
  standard model predictions with experimental results. In the event
  that deviations are observed, they provide a framework that allows to
  relate different experiments and cross-check possible theoretical
  interpretations. Being insufficient by themselves, e.g.\ because they
  are not renormalizable, parametrizations are expected to show the 
  directions to the correct underlying theory if deviations are
  observed. 
\item Models, sometimes even complete theories, provide specific
  frameworks that allow a consistent derivation of cross sections for
  conventional and new processes. Examples are the two-Higgs-doublet
  extension of the standard model, grand unified theories and, most
  importantly supersymmetry with or without $\rlap/\!R_p$-violation.
\end{itemize}
The following three examples attained most interest when the excess of
large-$Q^2$ events at HERA was made public \cite{altarelli,hsrr}. I will
try to point out some of the open questions worth to be studied in
future theoretical research.


\subsection{Contact interactions}

The contact interaction (CI) scenario relevant for HERA physics assumes
that 4-fermion processes are modified by additional terms in the
interaction Lagrangian of the form
\begin{equation}
{\cal L}_{\rm CI} = 
  \sum_{\small \begin{array}{c}i,k = L,R\\ q=u,d,\cdots
  \end{array}} \eta^q_{ik} 
  \frac{4\pi}{\left(\Lambda^q_{ik}\right)^2} \left(\bar{e}_i
    \gamma^{\mu} e_i\right) \left(\bar{q}_k \gamma_{\mu} q_k\right) .
\label{CI}
\end{equation}
Similar terms with 4-quark interactions would be relevant for new
physics searches at the Tevatron and 4-lepton terms would affect purely
leptonic interactions. In equation (\ref{CI}), as usual, only products
of vector or axial-vector currents are taken into account since limits
on scalar or tensor interactions are very stringent. Such terms are
motivated in many extensions of the standard model as effective
interactions after having integrated out new physics degrees of freedom
like heavy gauge bosons, leptoquarks and others, with masses beyond the
production threshold. The normalization with the factor $4\pi$ is
reminiscent of models which predict CI terms emerging from strong
interactions at a large mass scale $\Lambda$.  Beyond their meaning as
new physics effects, limits on the mass scale of contact interactions
serve as an important means to quantify the agreement of experimental
data with standard model predictions.

\begin{figure}[htbp]
\small
\begin{picture}(160,130)(-100,0)
\ArrowLine(50,120)(96,74)
\ArrowLine(104,66)(150,20)
\ArrowLine(50,20)(96,66)
\ArrowLine(104,74)(150,120)
\thicklines
\Line(96,66)(104,66)
\Line(104,66)(104,74)
\Line(104,74)(96,74)
\Line(96,74)(96,66)
\put(39,122){$e_i$}
\put(155,122){$\bar{e}_i$}
\put(39,18){$q_k$}
\put(155,18){$\bar{q}_k$}
\put(8,72){HERA}
\put(8,62){$ep \rightarrow eX$}
\put(54,67){$\Rightarrow$}
\put(90,130){LEP}
\put(69,120){$e^+e^- \rightarrow {\rm hadr}$}
\put(96.5,107){$\Downarrow$}
\put(80,12){Tevatron}
\put(70,2){$p\bar{p} \rightarrow \ell^+\ell^- X$}
\put(96.5,27){$\Uparrow$}
\end{picture}
\caption{\label{figCI}
  Schematic view of a contact interaction term.}
\end{figure}
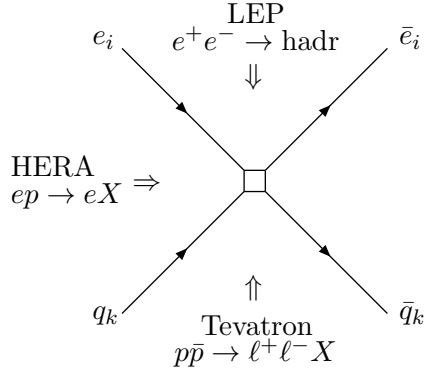

Equation \ref{CI} predicts modifications of cross sections for 4-fermion
processes in all channels as visualized in Fig.\ \ref{figCI}.  Both
enhancement or suppression are expected at the largest possible energies
if the CI mass scale is large, depending on the helicity structure of
the contact term and its sign $\eta^q_{ik}$. Due to the extremely high
experimental precision, also atomic parity violation experiments are
sensitive to parity-odd combinations of helicities. The important
advantage of the contact term approach is that it provides a framework
which can be applied to all presently running high-energy experiments.
The contact term approach relates predictions for DIS at HERA with
hadron production in electron positron annihilation and Drell-Yan
production at the Tevatron.

\begin{table}[ht]
\begin{center}
\begin{tabular}{llllllll}
\hline \hline \rule{0mm}{5mm}
&\multicolumn{2}{c}{\rule{0mm}{5mm}Drell-Yan}
&\multicolumn{2}{c}{\rule{0mm}{5mm}$ep \rightarrow eX$}
&\multicolumn{3}{c}{\rule{0mm}{5mm}$e^+e^- \rightarrow {\rm
    hadrons}$}\\[1mm] 
\cline{2-8} \rule{0mm}{5mm}
& CDF($ee,\mu\mu$) & D0($ee$) & H1 & ZEUS & Aleph & L3 & Opal \\[1mm]
\hline 
$VV+$ & 3.5 & 4.7 & 4.5 & 4.9 & 4.0 & 3.9 & 4.1 \\[1mm]
$VV-$ & 5.2 & 5.8 & 2.5 & 4.6 & 5.2 & 5.0 & 5.7 \\[1mm]
$AA+$ & 3.8 & 4.6 & 2.0 & 2.0 & 5.6 & 5.6 & 6.3 \\[1mm]
$AA-$ & 4.8 & 5.3 & 3.8 & 4.0 & 3.7 & 3.5 & 3.8 \\[1mm]
\hline \hline
\end{tabular}
\caption{Typical limits (in GeV) for the mass scale of CI terms from
  Tevatron, HERA and LEP for pure vector--vector and
  axial-vector--axial-vector type with positive or negative sign
  relative to the standard model interactions \protect\cite{cilimits}.}
\label{tabCI}
\end{center}
\end{table}

Table \ref{tabCI} gives a selection of recent limits on CI mass scales
as reported at the 1998 summer conferences (limits for other
combinations of helicities are available as well \cite{cilimits}). The
numbers in this table show that all present high-energy experiments have
achieved limits in a very similar mass range despite of their different
center-of-mass energies. Consequently, with a signal at HERA one should
expect visible effects at LEP2 and at the Tevatron. In the case of the
observation of deviations from the standard model predictions, the
combination of results obtained in different experiments and from
measurements with polarized beams \cite{CIpol} will be helpful to
identify the helicity structure of contact interaction terms.


\subsection{Leptoquarks}

Leptoquarks appear in extensions of the standard model involving
unification, technicolor, compositeness, or $R$-parity violating
supersymmetry. In addition to their couplings to the standard model
gauge bosons, leptoquarks have Yukawa-type couplings to lepton-quark
pairs which allow their resonant production in $ep$ scattering. Their
phenomenology in view of the observed excess of large-$x$, large-$Q^2$
events at HERA has been discussed extensively in the literature
(\cite{hsrr} and references therein). The generally adopted framework
described in Ref.\ \cite{BRW} is based on the assumption that the Yukawa
interactions of leptoquarks should have the following properties:
\begin{itemize}
\item renormalizability
\item $SU(3)\times SU(2)\times U(1)$ symmetry
\item conservation of baryon and lepton number
\item chirality of the couplings
\item couplings exist only to one fermion generation
\item no other interactions and/or particles exist
\end{itemize}
Dropping one of the first two of these assumptions would lead to severe
theoretical problems; the other properties are dictated by
phenomenology. One would certainly not like to give up assumption 3
since this avoids rapid proton decay. The chirality of couplings is
necessary in order to escape the very strong bounds from leptonic pion
decays and assumption 5 is a consequence of limits on FCNC processes.
The last assumption is made for simplicity only; it seems to be rather
unlikely than realistic.

These assumptions lead to a rather restricted set of allowed states and
their branching fractions $\beta$ to a charged lepton final state can
only be 1, 0.5, or 0. Those states which are interesting for HERA
phenomenology have $\beta = 1$ and are excluded by Tevatron bounds which
require masses above 242 GeV \cite{lqlimits}.

The leptoquark scenario might remain interesting if it is possible to
generalize the approach by relieving one or more of the above
assumptions \cite{hr}, notably the last one of the list. The Tevatron
mass bounds are avoided if it was possible to adjust the branching
ratios in the range $0.3 \lsim \beta \lsim 0.7$ \cite{altarelli,hr}.  In
Ref.\ \cite{babu} a scenario was proposed where two leptoquark states
show mixing induced by coupling them to the standard model Higgs boson.
Alternatively, interactions to new heavy fields might exist that, after
integrating them out, could lead to leptoquark Yukawa couplings as an
effective interaction \cite{agm}, bypassing this way renormalizability
as a condition since this is assumed to be restored at higher energies.
In the more systematic study of Ref.\ \cite{hr}, LQ couplings arise from
mixing of standard model fermions with new heavy fermions with
vector-like couplings and taking into account a coupling to the standard
model Higgs. Up to now, no attempt was made to study in a systematic way
the possibility of relieving the assumption that no intergenerational
couplings should exist (see however Ref.\ \cite{agm,ks}). The most
interesting extension of the generic leptoquark scenario is, however,
$R_p$-violating supersymmetry which is discussed in the next subsection.

 
\subsection{$R_p$-violating supersymmetry}

The Lagrangian of a supersymmetric version of the standard model may
contain a superpotential of the form
\begin{equation}
\begin{array}{lll}
W_{\not R_p} = & \phantom{+} \lambda_{ijk} L_i L_j E_k^c & 
  \hspace{10mm} \not\!\! L \\[1mm] 
& + \lambda'_{ijk} L_i Q_j D_k^c & 
  \hspace{10mm} \not\!\! L ~~~{\rm (includes LQ-like~couplings)} 
\\[1mm]
& + \lambda''_{ijk} U_i^c D_j^c D_k^c & \hspace{10mm} \not\!\! B
\end{array}
\label{rpviol}
\end{equation}
which violates lepton or baryon number conservation as indicated.
Imposing symmetry under $R$-parity (defined as $R_p = (-1)^{3B+L+2S}$)
forbids the presence of $W_{\not R_p}$.  The resulting phenomenology has
been searched for at all present high energy experiments and HERA may
set interesting limits which are complementary to those obtained at
Tevatron \cite{future}. Future experiments at the LHC will extend the
search limits for $R_p$-conserving supersymmetry considerably.

The present limits on the proton life-time do not forbid interactions of
the form $L_i Q_j D_k^c$ proportional to $\lambda'_{ijk}$ provided the
$\lambda''_{ijk}$ are chosen to be zero at the same time. This makes
squarks appear as leptoquarks which can be produced on resonance in
lepton-quark scattering.  In contrast to the generic leptoquark
scenarios described above, $R_p$-conserving decays of squarks lead to a
large number of interesting and distinct signatures (see Ref.\ 
\cite{dreiner} and references therein).  Characteristically one expects
multi-lepton and multi-jet final states.  The branching ratios can be
adjusted so as to avoid the strict mass limits from Tevatron.

Most of the analyses done so far assume that only one of the couplings
$\lambda'_{ijk}$ is non-zero and only one squark state is in reach. A
more general scenario with two light squark states has been considered
in Ref.\ \cite{kon1} where it was shown that
$\tilde{t}_L$--$\tilde{t}_R$ mixing would lead to a broader $x$
distribution than expected for single-resonance production.  The
possibility of having more than one $\lambda'_{ijk} \ne 0$ was noticed
in Ref.\ \cite{belyaev} and deserves more theoretical study.

$R_p$-violating supersymmetry has also played a role in the search for
explanations of the observation of a large number of events with an
isolated $\mu$ and missing transverse momentum \cite{h1muon}. Events of
this kind can originate from $W$ production followed by the decay $W
\rightarrow \mu \nu_{\mu}$; their observed number is, however, larger
than expected and their kinematical properties are atypical for $W$
production.  An explanation in terms of anomalous $WW\gamma$ couplings
additionally has to face limits from Tevatron and LEP2 and leaves the
question open why a similar excess of events is not seen in $e +
\rlap/p_T$ events.

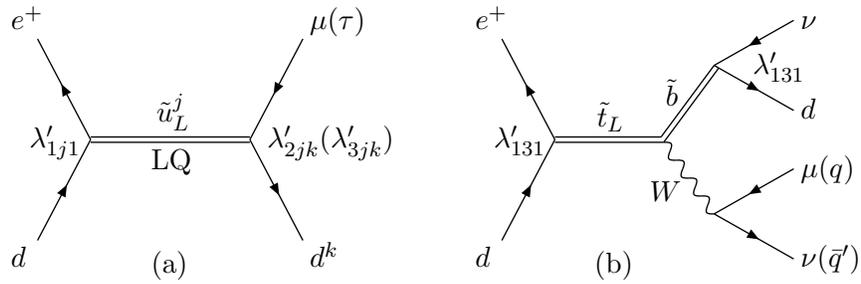
\begin{figure}[htbp]
\small
\begin{picture}(135,90)(-73,0)
\ArrowLine(10,12)(30,50)
\ArrowLine(30,50)(10,88)
\ArrowLine(110,88)(90,50)
\ArrowLine(90,50)(110,12)
\Line(30,51)(90,51)
\Line(30,49)(90,49)
\put(0,92){$e^+$}
\put(0,2){$d$}
\put(113,92){$\mu(\tau)$}
\put(113,2){$d^k$}
\put(55,57){$\tilde{u}_L^j$}
\put(53,39){LQ}
\put(7,47){$\lambda'_{1j1}$}
\put(97,47){$\lambda'_{2jk}$($\lambda'_{3jk}$)}
\put(53,0){(a)}
\end{picture}
\begin{picture}(135,90)(-110,0)
\ArrowLine(10,12)(30,50)
\ArrowLine(30,50)(10,88)
\Line(30,51)(70,51)
\Line(30,49)(71,49)
\Line(70,51)(90,78)
\Line(71,49)(91.74,77)
\Photon(71,49)(90,22){2}{4}
\ArrowLine(120,95)(90,78)
\ArrowLine(90,78)(120,61)
\ArrowLine(120,39)(90,22)
\ArrowLine(90,22)(120,5)
\put(0,92){$e^+$}
\put(0,2){$d$}
\put(123,92){$\nu$}
\put(123,58){$d$}
\put(123,2){$\nu$($\bar{q}'$)}
\put(123,36){$\mu$($q$)}
\put(46,55){$\tilde{t}_L$}
\put(7,47){$\lambda'_{131}$}
\put(105,75){$\lambda'_{131}$}
\put(72,63){$\tilde{b}$}
\put(66,27){$W$}
\put(45,0){(b)}
\end{picture}
\caption{\label{figrp1}
  Possible decays of squarks produced in $e^+d$ scattering with
  $R_p$-violating couplings leading to isolated $\mu$ + jet final
  states: (a) $\tilde{u}_L^j \rightarrow \mu d_k$ through
  $\lambda'_{2jk} \ne 0$; (b) $\tilde{t} \rightarrow \tilde{b} W$
  followed by $\tilde{b} \rightarrow \nu d$ via $\lambda'_{131} \ne 0$
  and $W \rightarrow \mu^+ \nu_{\mu}$ or $W \rightarrow$ 2 jets
  \protect\cite{kon2}.}
\end{figure}

\begin{figure}[htbp]
\begin{picture}(165,80)(-69,0)
\Line(10,49)(50,49)
\Line(10,51)(50,51)
\ArrowLine(50,50)(75,85)
\Photon(50,50)(90,50){2}{4}
\Line(50,50)(90,50)
\Photon(90,50)(115,15){2}{4}
\Line(90,50)(115,15)
\Photon(90,50)(130,50){2}{4}
\ArrowLine(155,85)(130,50)
\ArrowLine(130,50)(155,15)
\put(4,48){$\tilde{t}$}
\put(75,89){$b$}
\put(70,58){$\tilde{\chi}^+_1$}
\put(115,9){$\tilde{\chi}^0_1$}
\put(109,58){$W^+$}
\put(155,89){$\mu^+$}
\put(155,9){$\nu_{\mu}$}
\put(4,5){(a)}
\end{picture}
\begin{picture}(165,80)(-109,0)
\Line(10,49)(50,49)
\Line(10,51)(50,51)
\ArrowLine(50,50)(75,85)
\Photon(50,50)(90,50){2}{4}
\Line(50,50)(90,50)
\ArrowLine(90,50)(115,15)
\Line(90,49)(130,49)
\Line(90,51)(130,51)
\ArrowLine(155,85)(130,50)
\ArrowLine(130,50)(155,15)
\put(4,48){$\tilde{t}$}
\put(75,89){$b$}
\put(70,58){$\tilde{\chi}^+_1$}
\put(115,9){$\nu_{\ell}$}
\put(109,58){$\tilde{\ell}^+$}
\put(155,89){$\mu^+$}
\put(155,9){$\nu$}
\put(4,5){(b)}
\end{picture}
\caption{\label{figrp2}
  Possible decay chains of the stop leading to isolated muon + jet +
  missing $p_T$: $\tilde{t} \rightarrow b \tilde{\chi}_1^+$ followed by
  (a) $\tilde{\chi}_1^+ \rightarrow \tilde{\chi}_1^0 \mu^+ \nu_{\mu}$
  \protect\cite{kon3}; (b) $\tilde{\chi}_1^+ \rightarrow \nu_{\ell}
  \mu^+ \nu$ \protect\cite{muon5}.}
\end{figure}
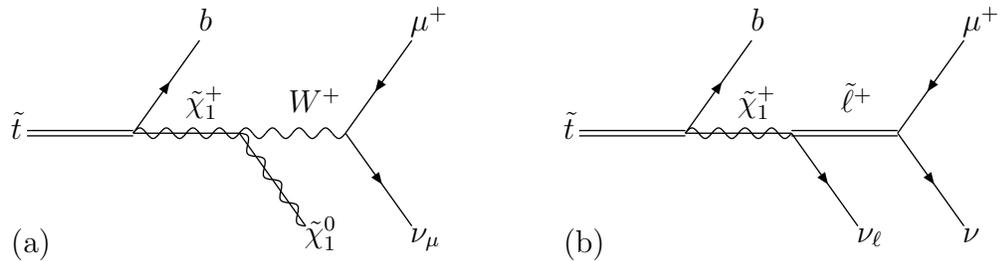

The observation of $\mu+\rlap/p_T$ events could find an explanation in
$R_p$-violating scenarios if it is assumed that a stop is produced
on-resonance at HERA.  Figures \ref{figrp1} and \ref{figrp2} show
examples for some of the possibilities. The process $ed \rightarrow
\tilde{t} \rightarrow \mu d^k$ (Fig.\ \ref{figrp1}a) which predicts
$\mu$ but no large $\rlap/p_T$ in the final state requires two different
non-zero $\lambda'$ couplings. The relevant product $\lambda'_{1j1}
\lambda'_{2jk}$ would induce flavor changing neutral currents and is
therefore limited to unreasonably small values for $1^{\rm st}$ and
$2^{\rm nd}$ generation quarks in the final state \cite{fcnclimits}.
The scenario shown in Fig.\ \ref{figrp1}b \cite{kon2} requires a
relatively light $b$ squark, $m_{\tilde{b}} \lsim 120$ GeV,and some
fine-tuning in order to avoid too large effects on $\Delta \rho$ in
electroweak precision measurements. It could be identified by the
simultaneous presence of multi-jet final states with $\rlap/p_T$ from
hadronic decays of the $W$.  Also the cascade decay shown in Fig.\ 
\ref{figrp2}a \cite{kon3} involving $R_p$-violation only for the
production of the $\tilde{t}$ resonance, not for its decay, seems
difficult to be achievable since it requires both a light chargino and a
long-lived neutralino. This, as well as the even more speculative
process shown in Fig.\ \ref{figrp2}b \cite{muon5} which requires
$R_p$-violation in the $L_i L_j E_k^c$ sector ($\lambda_{ijk} \ne 0$) as
well, can be checked from the event kinematics: assuming a value for the
mass of the decaying $\tilde{t}$, the recoil mass distribution must
cluster at a fixed value, the chargino mass. A more detailed discussion
of the $\mu+\rlap/p_T$ events and their possible theoretical origin can
be found in \cite{WG3muon}.

\section{Concluding remarks}

The search for new physics effects relies in many cases on trustworthy
predictions from the standard model, in particular when generic searches
look for ``interesting'' final states without having at hand a specific
model that tells the experimenter what and where to look for precisely.
New physics will always, if at all, show up at the frontier of the
experiments, i.e.\ at the largest energies or transverse momenta where
cross sections are smallest and experimental problems most severe. It is
therefore a mandatory though nontrivial task to combine the information
from as many as possible different experiments. In order to enhance the
statistical significance and reduce the probability that experimental
deficiencies lead to wrong interpretations, also experiments which did
not obtain the most stringent limits are important. The experiments at
HERA are therefore guaranteed to contribute to the search for new
physics.



\end{document}